# Competitive Effects of 2+ and 3+ Cations on DNA Compaction

C. Tongu*, T. Kenmotsu, Y. Yoshikawa, A. A. Zinchenko, N. Chen and K. Yoshikawa

**ABSTRACT**

By using single-DNA observation with fluorescence microscopy, we observed the effects of divalent and trivalent cations on the higher-order structure of giant DNA (T4 DNA with 166 kbp). It was found that divalent cations, such as Mg(2+) and Ca(2+), inhibit DNA compaction induced by a trivalent cation, spermidine (SPD(3+)). On the other hand, in the absence of SPD(3+), divalent cations cause the shrinkage of DNA. These experimental observations are inconsistent with the well-established Debye-Huckel scheme regarding the shielding effect of counter ions, which is given as the additivity of contributions of cations with different valences. We interpreted the competition between 2+ and 3+ cations in terms of the change in the translational entropy of the counter ions before and after the folding transition of DNA. For the compaction with SPD(3+), we considered the increase in translational entropy due to the ion-exchange of the intrinsic monovalent cations condensing on a highly-charged polyelectrolyte, double-stranded DNA, by the 3+ cations. In contrast, the presence of 2+ cation decreases the gain of entropy owe to the ion-exchange between monovalent and 3+ ions.

**KEYWORDS**: DNA condensation, higher-order structure of DNA, single-molecule observation, polyelectrolyte, counter ion condensation, polyamine

*E-mail:tongu@dmpl.doshisha.ac.jp

## INTRODUCTION

DNA is a polyelectrolyte with a high density of negative charge. Thus, an aqueous solution is generally a good solvent for DNA. Since the persistence length of duplex DNA is on the order of 50 nm (1, 2) and the distance between the base pairs is 0.34 nm (3), a DNA chain larger than several tens of kilo-base pairs (＞~10 kbp) behaves as a semi-flexible chain (4, 5), i.e., long DNA chains usually exhibit an elongated coiled state in an aqueous environment in vitro. On the other hand, genomic DNA molecules above the size of Mbp exist as condensed states in both prokaryotic and eukaryotic living cells. For example, in individual human cells, DNA chains with a total length on the order of a meter are stored in a small space about the size of a micrometer (3). Currently, the transition between the elongated coil and folded compact conformations of long DNA has attracted considerable interests from biologists, since this change in the higher-order structure of DNA is expected to be closely related to the mechanism of the self-regulation of replication and transcription in living cells (6). Various cationic chemical species, such as histone proteins, metal cations, and polyamines, have been experimentally shown to induce the compaction of long DNA chains (6-16). This conformational transition has also been studied theoretically (17-19). Among these chemical species, polyamines are widespread in both prokaryote and eukaryote cells, and exert various biological effects (6). In vitro experiments have shown that polyamines such as spermidine (3+) and spermine (4+) cause a large discrete phase transition of long DNA accompanied by a change in the effective molecular density on the order of $10^4$-$10^5$ (20, 21). On the other hand, the effect of divalent cations on the higher-order structure is less significant than that of polycations with a valence of $\geq 3$. Nevertheless, divalent cations also cause shrinkage of the DNA chain at relatively high concentrations of several to several tens of mM (22). In usual biological systems, polyamines with 3 or 4 positive charges and divalent metal cation, such as Mg(2+) or Ca(2+), are typical poly-cationic species. Thus, it may be of interest to clarify the effect of the coexistence of multivalent polyamines and divalent metal cations on the higher-order structure of a giant DNA molecule.

This report makes clear that divalent metal cations inhibit the ability of spermidine(3+) to cause compaction of giant DNA molecules, based on single-DNA observations using fluorescence microscopy. We propose a theoretical model to describe the competitive effect based on the change in translational entropy of the counter ions around polyelectrolytes.

## MATERIALS AND METHODS
### Materials
T4 GT7 DNA (166kbp) was purchased from Nippon Gene Co., LTD (Toyama, Japan). Spermidine-HCl, $MgCl_2$, $CaCl_2$ and NaCl were obtained from Nacalai Tesque (Kyoto, Japan).

The fluorescent dye Gelgreen was obtained from Biotium (CA, USA). The antioxidant 2-mercaptoethanol (2-ME) was purchased from Wako Pure Chemical Industries (Osaka, Japan).

**DNA conformation in solution as observed by fluorescence microscopy**

For fluorescence microscopic observations, T4 GT7 DNA at a final concentration of 0.1μM was dissolved in 1 mM Tris-HCl buffer solution at pH 7.5 with 5 μM Gelgreen and 4 %(v/v) 2-ME. For observation of the DNA conformation in solution, desired concentrations of SPD, $MgCl_2$, $CaCl_2$ and NaCl were added to the sample solutions. Fluorescence images of DNA molecules were captured by using a Axiovert 135 TV (Carl Zeiss, Jena, Germany) microscope equipped with an oil-immersed 100×objective lens and recorded on a DVD through an EBCCD camera (Hamamatsu Photonics, Hamamatsu, Japan). All observations were carried out at room temperature (24 °C).

**Observation of DNA conformation captured by atomic force microscopy (AFM)**

For AFM imaging by a scanning probe microscope (SPM; SPM-9700, Shimadzu Co., Kyoto, Japan), DNA samples were dropped onto a freshly cleaved mica surface. After incubation for 3-10 min at room temperature, the mica was rinsed with water and dried under nitrogen gas. The DNA concentration and buffer condition were the same as in the observation by fluorescence microscopy. All imaging was performed in air using the dynamic mode. The cantilever (OMCL-AC200TS-C3, Olympus) had a spring constant of 6.7-20.1 N/m. The scanning frequency was 0.4 Hz, and images were captured using the height mode in a 512×512 pixel format. The obtained images were plane-fitted and flattened by the computer program supplied with the imaging module before analysis.

**RESULTS**

Figure 1 shows snapshots of fluorescence microscopic observations of single T4 DNA molecules exhibiting Brownian motion in aqueous solution, together with quasi-three-dimensional images of fluorescence intensity and schematic representation of the DNA molecules as inferred from the fluorescence images. The addition of 10 mM $MgCl_2$ causes unfolding of the DNA compact state into an elongated coil state, as shown in Fig. 1(e). Further addition of $MgCl_2$ up to 30mM induces a transition back to the compact state, as shown in Fig. 1(f). Thus, it is clear that Mg(2+) has an antagonistic effect on the folding transition induced by SPD(3+).

Figure 2 shows histograms of the distributions of the long-axis length $L$ of T4 DNA depending on the concentrations of $MgCl_2$ and $CaCl_2$ in the absence and presence of SPD (3+).

Figure 2(a) shows the distribution of the long-axis length of T4 DNA molecules without a condensation agent such as Mg(2+), Ca(2+) or SPD(3+). The addition of SPD(3+) induced the transformation of T4 DNA molecules from an elongated coil state to a folded compact state (Fig. 2(b)).

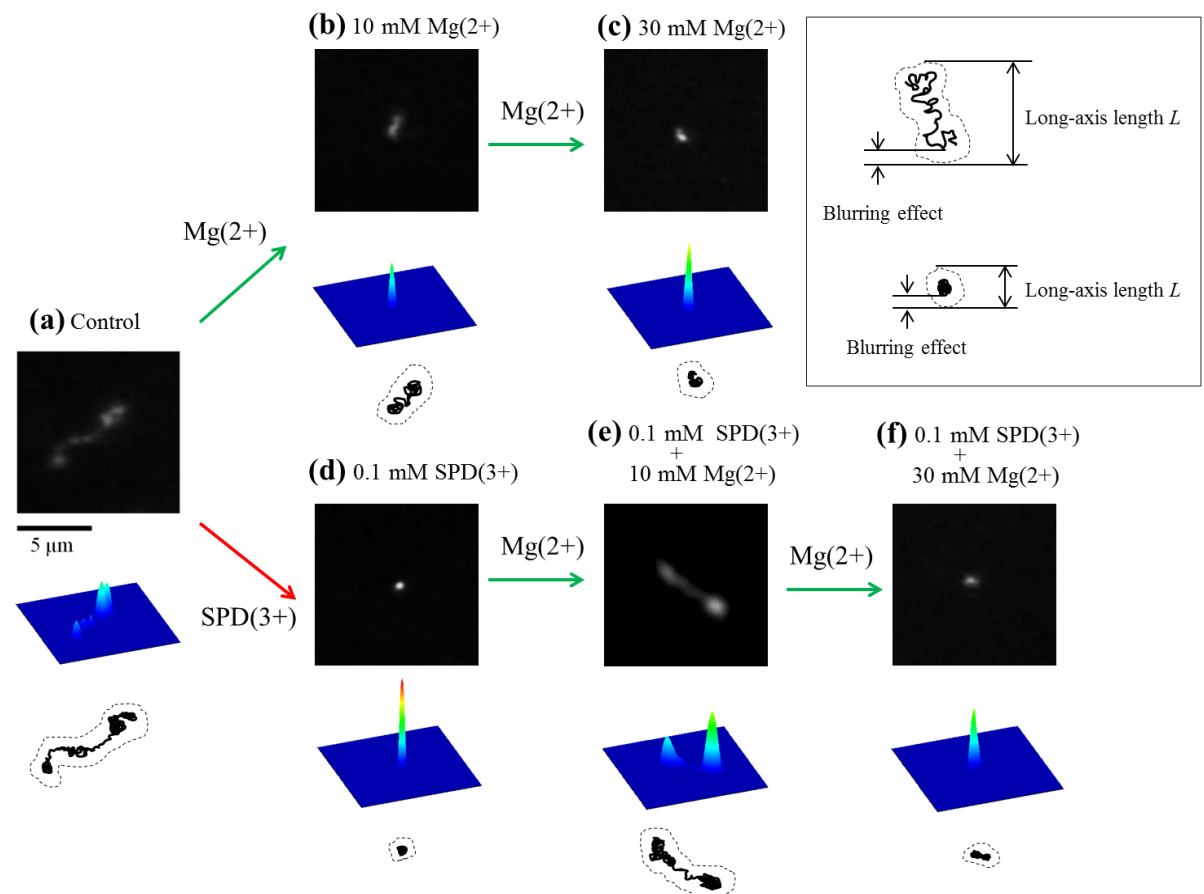

FIGURE 1 Conformational transition of DNA; b,c) with the addition of MgCl$_2$, d) with 0.1 mM SPD, and e,f) with further addition of MgCl$_2$ to the solution in d). Fluorescence images of T4 DNA together with quasi-three-dimensional representations of the fluorescence intensity, and schematic representation of DNA molecules as inferred from the fluorescence images.

The long-axis lengths deduced from single-T4 DNA observations contain blurring effect of around 0.3μm as shown schematically in Fig. 1. The average long-axis length of T4 DNA molecules is 2.7 μm for the control and 0.34 μm for 0.1 mM SPD(3+). In the absence of SPD (3+), Mg(2+) causes gradual shrinking of the DNA conformation, and at 30 mM the mean long-axis length becomes less than 1 μm, as shown in Fig. 2 (c). A similar shrinking effect is observed with the addition of Ca(2+); see Fig. 2(d). In contrast, in the presence of SPD(3+), Mg(2+) induces the unfolding of compact DNA; at 10 mM, the average value of $L$ becomes ca. 2.0 μm, as shown in Fig. 2(e). With a further increase in the Mg(2+) concentration, DNA tends to shrink slightly. Ca(2+) has a similar unfolding effect on compact DNA, as shown in Fig. 2(f).

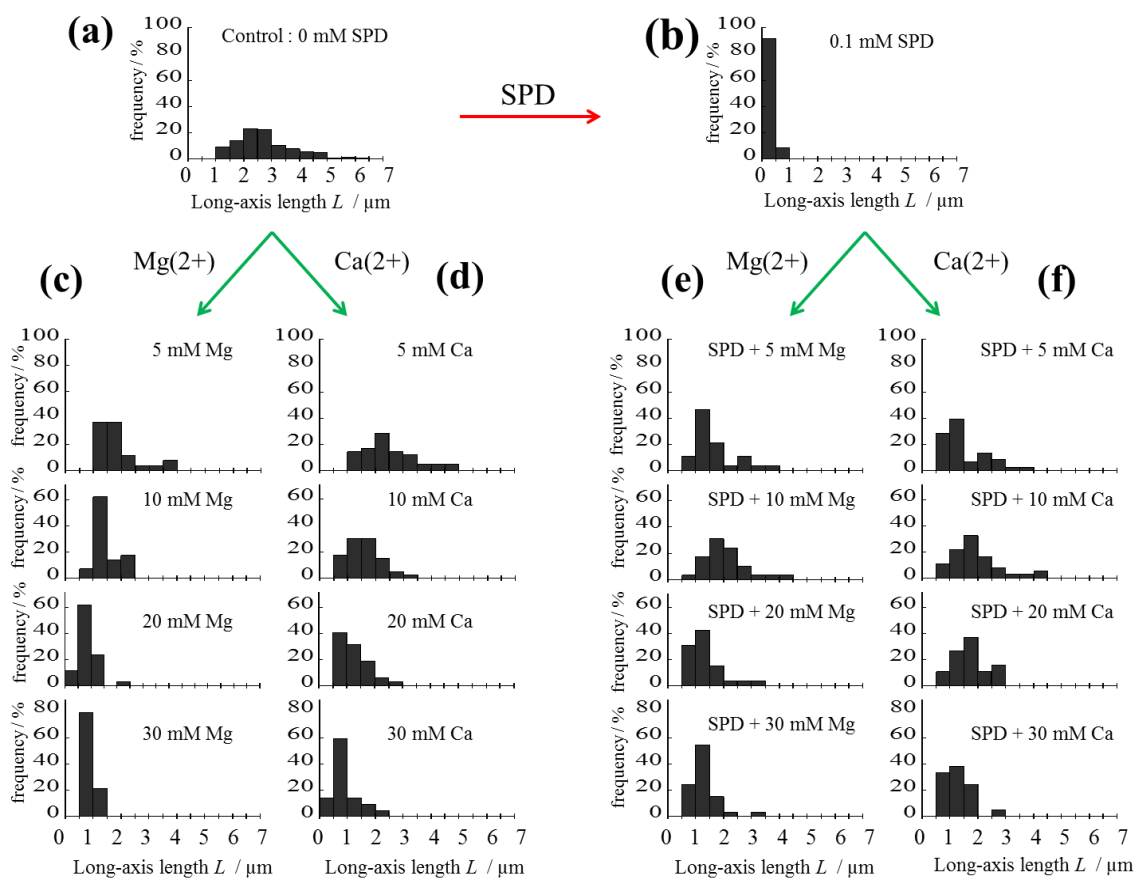

FIGURE 2 Distributions of long-axis length of T4 DNA at various concentrations of MgCl$_2$ and CaCl$_2$ containing 0 and 0.1 mM SPD.

To compare the effects of monovalent and divalent cations, we also monitored the DNA conformation by changing the concentration of NaCl, both in the absence and presence of SPD(3+). Figure 3(a) shows that $L$ decreases slightly with an increase in the NaCl concentration. This rather weak shrinking effect by a monovalent cation is attributable to a decrease in the persistence length, which was discussed previously from both experimental and theoretical perspectives (23). Furthermore, Na(+) inhibits the DNA compaction induced by SPD(3+), as shown in Fig. 3(b) in a good agreement with earlier studies (6, 24, 25). It is now clear that divalent cations, Mg(2+) and Ca(2+), as well as a monovalent cation, Na(+), inhibit spermidine-induced DNA compaction. The difference between divalent and monovalent cations is that the former condense DNA, whereas the latter have almost no apparent condensing effect.

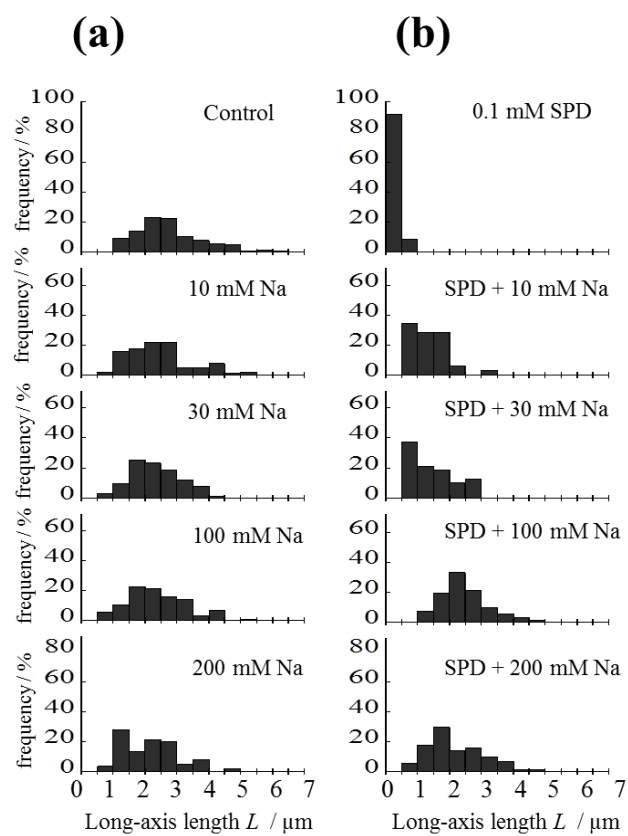

FIGURE 3 Distributions of long-axis length of T4 DNA at various concentrations of NaCl containing 0 and 0.1 mM SPD.

To grasp the general trend of the effect of coexisting cations with different valences, as depicted in the histograms in Figs. 2 and 3, in Fig. 4 we plotted the change in the mean value of the DNA long-axis length $L$ depending on the concentrations of $MgCl_2$, $CaCl_2$ and NaCl. Figure 4(a) and (b) indicate the gradual shrinkage of DNA by divalent cations in the absence of SPD(3+). In contrast, with 0.1mM SPD(3+), the compact DNA unfolds into an elongated coil state at concentrations of around 5-10 mM Mg(2+) and Ca(2+), as shown by the blue dotted lines in Fig. 4(a) and (b), respectively. With a further increase in the concentration of divalent cations, the long-axis length $L$ becomes smaller, but still reflects an elongated conformation up to a divalent cation concentration of 30 mM. In both cases, DNA molecules show greater folding above a concentration of 10 mM. The unfolding of compact DNA is also observed with the addition of monovalent Na(+), as shown in Fig. 4(c), where the potentiality upon unfolding is much weaker than that of divalent cations; i.e. a one-order-of-magnitude greater Na(+) concentration is necessary to unfold the DNA.

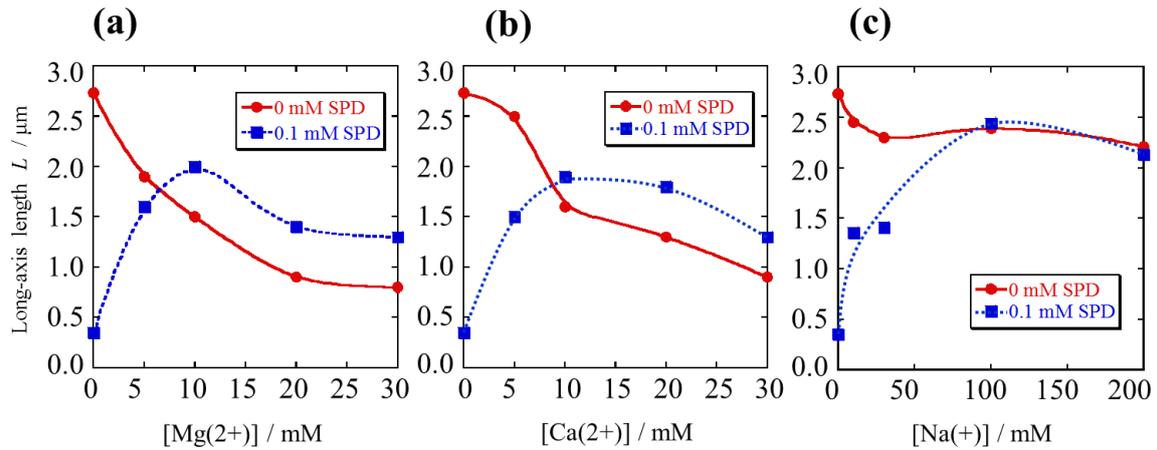

FIGURE 4 Mean values of the long-axis length L in the absence and presence of SPD(3+), depending on the concentrations of $MgCl_2$, $CaCl_2$ and NaCl.

The AFM images in Fig. 5 show the detailed DNA morphology just before compaction by SPD(3+) and shrinkage by Mg(2+). With 0.1 mM SPD(3+), a looping structure consisting of essentially straight chains with a parallel alignment appeared, as shown in Fig. 5(a) (26). With 20 mM $MgCl_2$, a cross-linked structure was observed (Fig. 5(b)). This cross-linked structure corresponds well to that in a previous study (27). These observations reflect the difference in the change in the DNA conformation caused by SPD(3+) and Mg(2+). Fig. 5(b-c) reveals that the number of cross-links tends to increase with the increase of Mg(2+) concentration. Even with an increase in the $MgCl_2$ concentration, a looping structure similar to that with SPD(3+) did not appear in our observation (Fig. 5(c)). This indicates that Mg(2+) causes shrinkage of a long DNA molecule, being much different of the effect of SPD(3+) to induce DNA compaction. Here, it is to be mentioned that the AFM images were obtained for the DNA absorbed onto a solid substrate, whereas the fluorescence microscopic images were observed for the DNA existing in bulk solution. Based on the information for the bulk solution, we may perform rather reliable interpretation on the detailed conformation observed for the absorbed state on a substrate.

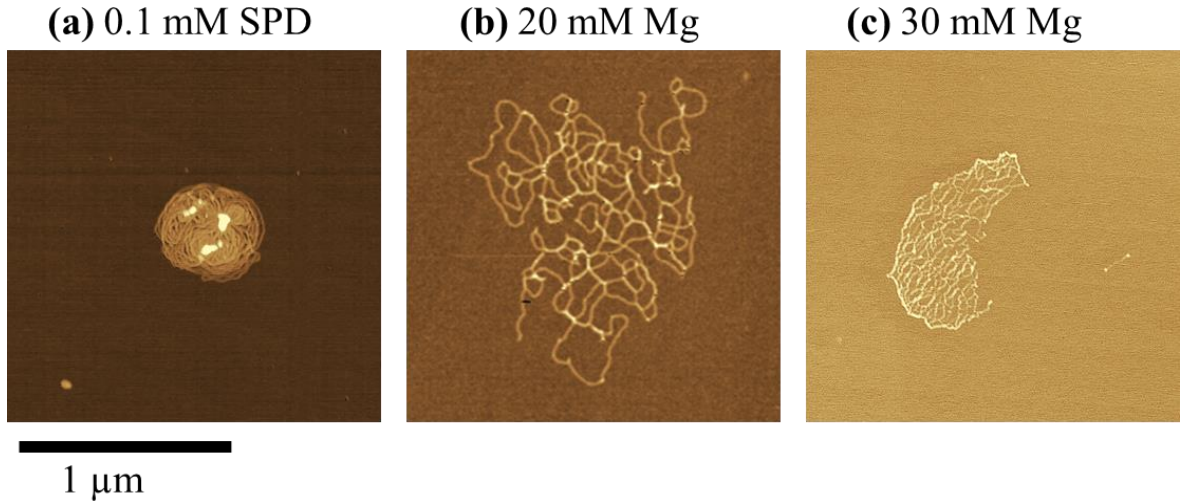

FIGURE 5 AFM images of T4 DNA under conditions of a) 0.1 mM SPD , b) 20 mM MgCl$_2$ and c) 30 mM MgCl$_2$. The images show high parallel arrangement of DNA segments in a), whereas segments tend to cross each other with almost no formation of a parallel arrangement in b-c).

**DISCUSSION**

The experimental results indicate that Ca(2+) and Mg(2+) ions indicate an antagonistic effect on the folding transition of DNA molecules. In general, the shielding efficiency of electronically charged species is interpreted as the sum of the effects of the surrounding counter ions with different valences: Debye length $\lambda_D \sim I^{-1/2}$, where ionic strength $I \sim \sum c_i Z_i^2$ (28). On the other hand, the present experimental results demonstrate that divalent cations inhibit SPD(3+)-induced compaction. The breakdown of the classical Debye picture of DNA compaction is attributable to the strong correlation effect of negative charges aligned along the double-stranded DNA, which is a highly charged polyelectrolyte chain with a relatively large persistence length (~50nm). Thus, we have to take into account the effect of counter ion condensation (29). In a usual aqueous environment with only monovalent counter ions, around 70% of the total negative charge of DNA is neutralized by monovalent cations (29, 30). As an additional important factor, we have to consider ion-exchange between monovalent and multivalent cations in the transition of giant DNA from an elongated to a compact state (20). This effect contributes to stabilization of the compact state through an increase in translational entropy due to the release of monovalent condensed counter ions into the bulk solution, through an ion-exchange mechanism. When this increase in translational entropy becomes significant, the free energy penalty on the conformational entropy of the compact state, together with a decrease in translational entropy to fully neutralize the negative charge of the DNA chain, can be compensated. This scenario for the change in free energy well explains the inhibitory effect of monovalent cations on compaction, as shown in Fig. 4(c). In the following discussion, we consider the change in free energy upon the folding transition, mainly focusing on the effect of the translational entropy of coexisting counter ions with different valences.

We consider the total free energy $F_{total}$ per single giant DNA molecule.

$$F_{total} = F_{elas} + F_{trans}, \tag{1}$$

where $F_{elas}$ and $F_{trans}$ are the elastic energy of the DNA chain and the free energy contribution from the translational entropy of a counter ion, respectively. The elastic free energy of DNA is given as follows by adopting the parameter $\alpha = R/R_0$, where $R$ and $R_0$ are the radius of the actual state and the ideal state of a DNA molecule, respectively (4, 20, 31):

$$F_{elas} = \left(\frac{3}{2}\right) k_B T \left(\alpha^2 + \alpha^{-2}\right), \tag{2}$$

where $k_B$ is the Boltzmann constant. $T$ is temperature, which was set to 300 K in the present work. We discuss the change in free energy that accompanies the folding transition between the coil and compact states:

$$\Delta F_{total} = \Delta F_{elas} + \Delta F_{trans}, \tag{3}$$

where $\Delta F_{elas} > 0$. We evaluate the free energy due to translational entropy in the coil state with only a monovalent cation as the counter ion at a concentration of $C_1$. The total number of phosphate groups (negative charge of DNA) in a single DNA molecule is $\Gamma_0$. The degree of counter ion condensation with respect to the number of phosphate groups is given as $p$, with monovalent cations in aqueous solution under room temperature, which is deduced from the Bjerrum length (29, 30). Thus, the free energy cost for counter ion condensation in the coil state is $p\Gamma_0 \ln C_1$, by adopting the unit of energy $k_B T$. Next, we estimate the free energy cost due to translational entropy in the folding transition to the compact state.

For simplification, in the following discussion we adopt the approximation of complete charge neutralization in a compact state (32). First, we consider the change in free energy upon the folding transition for the coexistence of monovalent and trivalent cations, $C_1$ and $C_3$, respectively. We estimate the change in free energy $\Delta F_{trans}^{1,3}$ from the contribution of translational entropy, for a simplified case when a DNA chain is fully neutralized by $\Gamma_0/3$ trivalent cations, by releasing $p\Gamma_0$ monovalent cations:

$$\Delta F_{trans}^{1,3} = -\frac{1}{3}\Gamma_0 \ln C_3 + p\Gamma_0 \ln C_1 . \tag{4}$$

Next, we consider the free energy cost due to translational entropy to cause the folding transition, under the coexistence of monovalent, divalent and trivalent cations, with concentrations of $C_1$, $C_2$ and $C_3$, respectively. Again, for simplicity, we consider the fully charge-neutralized state:

$$\Delta F_{trans}^{1,2,3} = -\Gamma_3 \ln C_3 - \Gamma_2 \ln C_2 + p\Gamma_0 \ln C_1 , \tag{5}$$

where $\Gamma_0 = 3\Gamma_3 + 2\Gamma_2$.

We may roughly consider the relationship between $\Gamma_2$ and $\Gamma_3$ by adopting a simple approximation as in the usual Debye-Huckel theory, i.e., linear approximation of the Poisson-Boltzmann equation:

$$\frac{\Gamma_3}{\Gamma_2} = \frac{9C_3}{4C_2} . \tag{6}$$

Then,

$$\Gamma_2 = \frac{4C_2}{27C_3 + 8C_2} \Gamma_0 . \tag{7}$$

Using the above relationships,

$$\Delta F_{trans}^{1,2,3} = -\frac{4C_2}{27C_3 + 8C_2}\left(\frac{9C_3}{4C_2}\ln C_3 + \ln C_2\right)\Gamma_0 + p\Gamma_0 \ln C_1 , \tag{8}$$

$$\Delta F_{trans}^{1,2} = -\frac{\Gamma_0}{2}\ln C_2 + p\Gamma_0 \ln C_1 . \tag{9}$$

Based on these arguments, DNA has an elongated coil state when the change in free energy calculated with eqs. (8) and (9) is positive. Meanwhile, the compaction of DNA occurs when the change in free energy is negative. If we assume that DNA exists in either a coil or globule state, where the difference in free energy for DNA is described by eq. (8) or (9), the probability $p$ of the transformation from the coil to globule states obeys the Boltzmann distribution. The

probability is given as

$$p_B = \exp\left(-\frac{\Delta F}{k_B T}\right), \tag{10}$$

where $\Delta F$ is the difference in free energy for DNA between the two states. Based on the present discussion, we may calculate the average long-axis length $L^{calc}$ of DNA as

$$L^{calc} = p_B L^{calc}_{coil} + (1 - p_B) L^{calc}_{globule}, \tag{11}$$

where $L^{calc}_{coil}$ and $L^{calc}_{globule}$ are the long-axis lengths of the coil and globule states, respectively.

Based on the experimental observations in Figs. 2, 3 and 4, we adopt 3.0 μm and 0.3 μm as the long-axis lengths for the coil and globule states, respectively. The Tris-HCl buffer solution which we used contains monovalent cations at a concentration of 0.1 mM. For simplicity, we considered the monovalent cations as the background, and the last term on the right-hand side in eqs. (8) and (9) was set to -7.7 $k_B T$ with a temperature of 300 K. Figure 6 shows the normalized long-axis length, $l$, calculated by $L^{calc}/L^{calc}_{coil}$ for a T4 DNA molecule at various concentrations of MgCl$_2$ with 0 and 0.1 mM SPD(3+) at 300 K. The concentrations of the monovalent and trivalent cations $C_1$ and $C_3$ were set to 0.1 and 0.01, respectively. The calculated long-axis length decreased with an increase in the concentration of divalent cation Mg(2+) in the absence of trivalent cation (SPD(3+)). The present theoretical discussion can explain the observed competitive effect under the coexistence of divalent and trivalent cations. It is important to note that the present theoretical model predicts how divalent cations contribute to the conformational transition of T4 DNA in the absence or presence of trivalent cations in solution by taking into account both counter ion condensation and the cost in translational entropy due to ion-exchange under the following assumption: as the number of phosphate groups of DNA, we set $\Gamma_0 = 10$ for the case in the presence of only Mg(2+) and $\Gamma_0 = 100$ for the case in the presence of SPD(3+), to describe the competitive effects of divalent and trivalent cations. The number of phosphate groups $\Gamma_0$ for SPD(3+) is an order of magnitude larger than that for Mg(2+). Here, we introduced a relatively large difference in the parameter $\Gamma_0$ between the effects of trivalent and divalent cations, due to the observed difference in the degree of DNA folding. Actually, as shown in Fig. 4 (a) and (b), the mean value of the long-axis length $L$ of shrunken DNA with 30 mM Mg(2+), and also with 30 mM Ca(2+), is 0.8 – 0.9 μm, in the absence of SPD(3+). In contrast, the mean value for the compact state with 0.1 mM SPD(3+) is around 0.4 μm, as shown in Fig. 2 (b). This large difference in the folded state indicates a relatively large difference in their higher-order structure. It is well known that SPD(3+) induces a tightly packed state for DNA, through a large discrete conformational transition. In contrast, as shown in the histograms in Fig. 2, divalent cations induce a gradual change in the DNA conformation. Based on this difference in the effects of divalent and trivalent cations, it is expected that cooperativity in the conformational transition is much lower for divalent metal cations than for SPD(3+). This argument is supported by a detailed inspection of the morphological changes by AFM, as shown in Fig. 5, where the compact conformation induced by SPD(3+) exhibits the parallel alignment of DNA segments accompanied by a seemingly uniform arrangement over the entire single T4 DNA. On the other hand, a large number of random crossings are found for the shrunken structure induced by Mg(2+). The difference in the parameterization on $\Gamma_0$ may thus be validated by considering this difference in the cooperativity of the folding transition between divalent and trivalent cations.

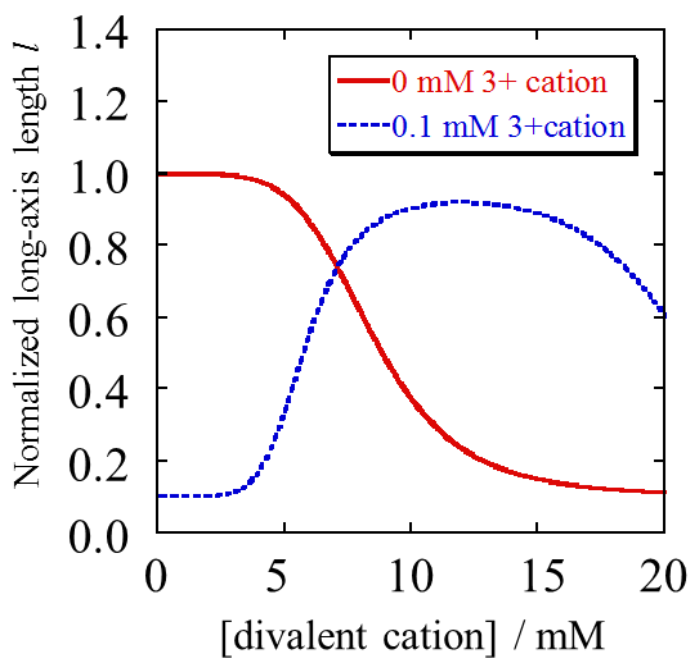

FIGURE 6 Calculated long-axis length of DNA molecules in the absence and presence of SPD(3+) depending on the concentration of divalent cation. For details of the theoretical treatment, see the text.

## CONCLUSION

We studied the change in the higher-order structure of giant DNA (T4 DNA with 166kbp) with divalent metal cations, Mg(2+) and Ca(2+), and with SPD(3+). While both the divalent and trivalent cations induce the folding of DNA molecules, the trivalent cation induces compaction with much greater potency. Interestingly, a divalent cation inhibits the folding transition of DNA induced by SPD(3+). We proposed a theoretical model to describe this competitive effect by taking into account the ion-exchange between monovalent cations of a DNA chain and multivalent cations surrounding DNA based on the argument of counter ion condensation. The correlation effect among negative charges along the double-stranded DNA chain and condensed counter ions play the essential role for the competitive or antagonistic effects of divalent and trivalent cations. The experimental results and theoretical arguments in the present work are expected to provide a fundamental understanding of the physic-chemical characteristics of the higher-order structure of giant genomic DNA molecules, including their interactions with histone and other various charged species in living cellular environments.

## AUTHOR CONTRIBUTIONS

## ACKNOWLEDGEMENTS

This work was partly supported by KAKENHI, Grants-in-Aid for Scientific Research, 15H02121 and 25103012.